\documentclass[11pt,twoside]{article}
\usepackage{newpasp}
\usepackage{epsf}
\usepackage{graphics}

\newcommand{\chandra}{{\em Chandra}}
\newcommand{\rosat}{{\em ROSAT}}

\begin{document}

\title {Synchrotron Radiation from Outer Space and the Chandra X-Ray Observatory\\
}

\author{Martin C. Weisskopf}
\affil{NASA/MSFC, VP62, MSFC AL 35812}

\begin{abstract}
The universe provides numerous extremely interesting astrophysical sources of synchrotron X radiation.
The \chandra\ X-ray Observatory and other X-ray missions provide powerful probes of these and other cosmic X-ray sources.
\chandra\ is the X-ray component of NASA's Great Observatory Program which also includes the Hubble Space telescope, the Spitzer Infrared Telescope Facility, and the now defunct Compton Gamma-Ray Observatory.  
The \chandra\ X-Ray Observatory provides the best angular resolution (sub-arcsecond) of any previous, current, or planned (for the foreseeable near future) space-based X-ray instrumentation. 
We present here a brief overview of the technical capability of this X-Ray observatory and some of the remarkable discoveries involving cosmic synchrotron sources.
\end{abstract}

\section{Technical Description of the \chandra\ X-Ray Observatory}

In this section we present an overview of the technical capability of the Observatory.
For more details, please see Weisskopf et al. (2003)\footnote{See also http://asc.harvard.edu/proposer/POG/index.html}.

\subsection{The Orbit and the Spacecraft}

\begin{figure}
\begin{center} 
\epsfysize=7.0cm
\epsfbox{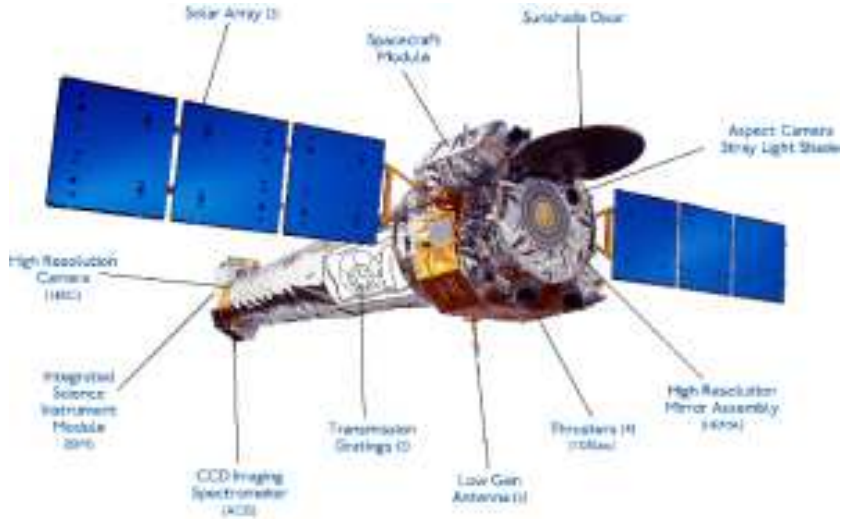} 
\caption{Drawing of the \chandra\ X-Ray Observatory with major components labeled. The spacecraft has a 4800-kg mass without propellant, 13-m length and 19.5-m wingspan. The separation between the telescope's optical node and the detector focal plane is 10 m. Courtesy NASA \label{f:spacecraft}}
\end{center}
\end{figure}

The \chandra\ X-Ray Observatory began its journey into space on July 23, 1999 using the Space Shuttle Columbia for the initial ascent. 
An upper stage and integral propulsion then placed the observatory into an elliptical orbit with an initial apogee of 140,000 km (1/3 of the distance to the moon) and an initial perigee 10,000 km (about two earth radii). 
In this orbit, the satellite (Figure~\ref{f:spacecraft}\footnote{Pictures that are publicly available at the \chandra\ web site at http://chandra.harvard.edu have credits labeled ``Courtesy of NASA''}) is above the radiation belts for more than 75\% of the 63.5-hour orbital period, allowing for uninterrupted observations up to more than 2 days.

\subsection{The Optics}

The heart of the Observatory is its telescope made of four concentric, precision-figured, superpolished, grazing-incidence Wolter (type-1) telescopes.
The design is similar to that used on the previous X-Ray observatories --- Einstein and \rosat\ ---  but of much higher quality, larger diameter, and longer focal length.
The telescope's on-axis point spread function has a full-width at half-maximum of about 0.2 arcsec and a half-power diameter less than 1 arcsec. 
There is a relatively mild dependence on energy, resulting from diffractive scattering by the optic's low ($< 3$-\AA) surface microroughness.

\subsection{The Gratings}

Just behind the telescope and along the path to the focal plane are 2 objective transmission gratings (OTGs) --- the Low-Energy Transmission Grating (LETG) and the High-Energy Transmission Grating (HETG).
Positioning mechanisms are used to insert either OTG (or none) into the converging beam when commanded.

The Space Research Organization of the Netherlands (SRON) and the Max-Planck-Instit\"ut f\"ur extraterrestrische Physik (MPE) provided the LETG.
With free-standing gold bars of about 991-nm period, the LETG allows for
high-resolution spectroscopy ($E/\Delta E$ $>1000$) between 175 and 80 \AA\/
(0.07 -- 0.15 keV) in first order and moderate resolving power at shorter wavelengths.  
The LETG~wavelength range accessible with the HRC-S as the detector is
 175 -- 1.2 \AA~(0.07 -- 10 keV).

MIT provided the HETG.
The HETG uses 2 types of grating~--- the Medium-Energy Gratings (MEG),
which, when inserted are behind the X-ray telescope's 2 outermost shells, and the High-Energy Gratings (HEG), the 2 innermost shells.
The HETG uses polyimide-supported gold bars of 400-nm and 200-nm periods and provides high-resolution spectroscopy from 0.4 to 4 keV (MEG, 30 to 3
\AA) and from 0.8 to 8 keV (HEG, 15 to 1.5 \AA) with ACIS-S as the detector.

\subsection{The Focal Plane Cameras \label{s:cameras}}

The Pennsylvania State University (PSU) and working with the Massachusetts Institute of Technology (MIT) designed and fabricated the Advanced CCD Imaging Spectrometer (ACIS), with CCDs produced by MIT's Lincoln Laboratory.
Made of a 2-by-2 array of large-format, front-illuminated (FI), 2.5-cm-square
CCDs, ACIS-I provides high-resolution spectrometric imaging over a
17-arcmin-square field of view.
ACIS-S, a 6-by-1 array of 4 FI CCDs and two back-illuminated (BI) CCDs, mounted
along the dispersion direction of \chandra's transmission gratings serves as the primary read-out detector for the HETG.
In addition, using the one BI CCD which can be placed at the aimpoint of the telescope, ACIS-S also provides high-resolution spectrometric imaging extending to lower energies
but over a smaller (8-arcmin-square) field than ACIS-I.
The spatial resolution for imaging with ACIS~is limited by the physical size of the CCD~pixels (24.0 ${\mu}$m square $\approx$0.492 arcsec). 
Each CCD has 1M pixels.

The Smithsonian Astrophysical Observatory (SAO), provided the other focal-plane imager --- the High Resolution Camera (HRC).
Made of a single 10-cm-square microchannel plate (MCP), the HRC-I provides
high-resolution imaging over a 30-arcmin-square field of view.
The HRC-S, 3 rectangular MCP segments (3-cm $\times$ 10-cm each) mounted end-to-end along the grating dispersion direction, serves as the primary read-out detector for the LETG.
A mechanical translation stage is moves the appropriate detector (ACIS-I, ACIS-S, HRC-I, HRC-S) to the focal point.

\section{Cosmic X-ray Synchrotron Sources \label{s:sources}} 

Synchrotron sources --- both laboratory and cosmic ---  play an important role in the \chandra\ project.
Prior to launch, calibration of elements of the Observatory utilized a number of ground-based synchrotrons including the BESSY synchrotron light source in Berlin, the ALS at Berkeley, and the NSLS at Brookhaven. 
Post launch, of course, \chandra\ observations have studied a large number of astronomical sources of synchrotron radiation.

\subsection{The Crab Nebula \label{ss:crab}}

\chandra's first observations of cosmic X rays occurred on August 12, 1999. 
Several of the early targets were remnants of a supernovae --- the aftermath of the explosion of a star. 
Supernovae are obvious targets for an X-ray observatory for a number of reasons.
Amongst these are the X-rays emitted when the shock waves and debris interact with the interstellar medium, producing temperatures of millions of degrees.
Depending upon the mass of the progenitor star, a compact object --- white dwarf, neutron star, or black hole may form. 
These compact stellar objects emit X-rays through a variety of mechanisms.
In the case of the black hole, the X rays originate from the nearby vicinity in that no light within the event horizon escapes. 

The Crab Nebula is the remnant of a supernova explosion observed by the Chinese in 1054 A.D. 
It is the most famous of this class of objects.
(For an excellent summary of the scientific history of the Crab Nebula and its pulsar see the book by Manchester and Taylor, 1977.)
Discoveries made from observations of the Crab have played an important role in the history of astronomy and astrophysics. 
In 1731 John Bevis, an English physicist and amateur astronomer, discovered the faded nebulosity. 
In 1758, Charles Messier included this object in his famous catalog as M1.
In 1844 the astronomer Lord Rosse noted the nebular emission and associated the shape that he saw with that of a crab, hence the name.
It is interesting that Hubble was the first to associate the nebula with the event of 1054.
The extended optical emission is partially line-emitting filaments, but the bulk is continuum emission.
In 1952, a Russian theorist, Shklovski, suggested that the optical continuum resulted from the synchrotron process.
This hypothesis was dramatically confirmed by the subsequent measurement of strong optical polarization.

In 1948, the Crab Nebula became the first radio source identified with an extra-solar system optical object.
In the early sixties, with the advent of X-ray astronomy, the Crab Nebula became one of the first identified X-ray sources.
The size of the nebula decreases with increasing energy, and this also points to synchrotron radiation.
Radio astronomers also found a compact radio source at the center of the nebula in 1964, and discovered radio pulsations in 1968. 
Optical and X-ray observations later also detected this pulsation ---- a rotating neutron star.
Measurement of the strong linear polarization of the nebular X-ray emission, together with the agreement of the position angle with that found in the visible, confirmed the synchrotron origin of the nebular emission from the radio to the X-ray bands. 

An important astrophysical question was how to power the nebula in that the  synchrotron lifetime, especially for X-ray emission, is much less than the lifetime of the nebula.
Measurement of the properties of the pulsar, specifically the loss of rotational energy with time, provided the qualitative answer.
The pulsar loses rotational energy and, somehow, converts it to relativistic particle energies which then synchrotron radiate in the nebular magnetic field. 

\begin{figure}
\begin{center} 
\epsfysize=6.5cm
\epsfbox{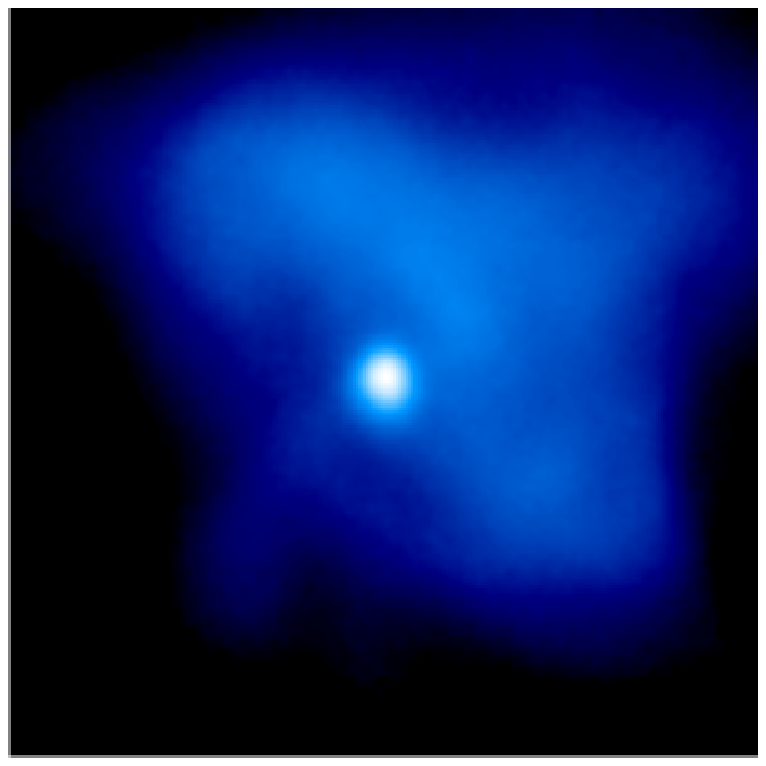} 
\epsfysize=6.5cm
\epsfbox{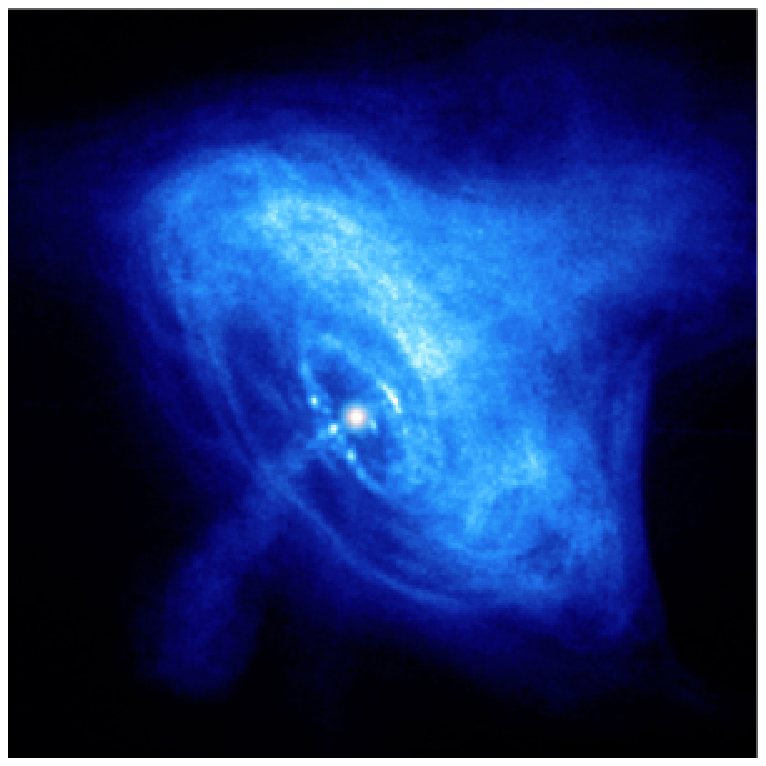} 
\caption{Observations of the Crab Nebula with two different X-ray observatories \rosat\ (to the left) and  \chandra\ image (to the right). Each image is 2.24' on a side. The \rosat\ image was provided by V. Zavlin. The \chandra\ image is courtesy of NASA.
\label{f:crab}}
\end{center}
\end{figure}

Despite the impressive history of accomplishments, \chandra\ provided another major discovery concerning the details of this process.
Figure~\ref{f:crab} shows two X-ray images the Crab Nebula and its central pulsar. 
The image to the left is from \rosat, a predecessor to \chandra, with about 4-arcsec angular resolution.
The image to the right is the \chandra\ image, which reveals the bright inner elliptical ring, showing the first direct observation of the shock front where the wind of particles from the pulsar begins to radiate in X rays via the synchrotron process (Weisskopf et al. 2000).
These \chandra\ observations have also provided new information such as the clearly resolved X-ray torus, jet and counterjet; the suggestion of a hollow-tube structure for the torus; X-ray knots along the inner ring and (perhaps) along the inward extension of the X-ray jet, etc.
Time-lapsed \chandra\ images (Hester et al.\ 2002) revealed nebular features moving at velocities about 1/2 the speed of light.

\subsection{A jet from a low-mass X-ray binary containing a Black Hole      \label{ss:bhlmxb}}

\begin{figure}
\begin{center} 
\epsfysize=7.0cm
\epsfbox{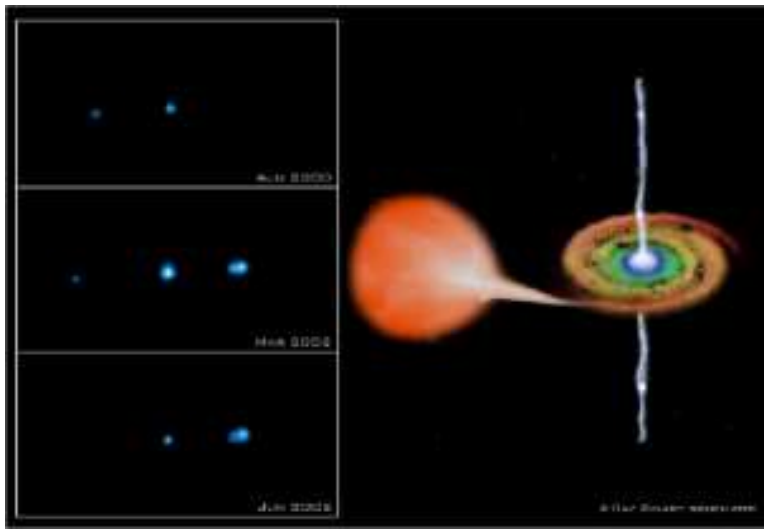} 
\caption{The three figures to the left are each 1.64' $\times$ 0.82' and were taken at different times. The emission from the central source originates in from an accretion disc formed outside the event horizon, through mass transfer from a normal companion, as illustrated in the cartoon to the right. Courtesy NASA.
\label{f:xtej1550}}
\end{center}
\end{figure}

The \chandra\ image of the Crab shows X-ray-emitting jets in but one astronomical context.
The jet phenomenon now appears in many astronomical settings.
A series of \chandra\ images allowed Corbel et al. (2002) to trace the evolution of the X-ray jets produced by a black hole in a binary star system (XTE J1550-564). 
Figure~\ref{f:xtej1550} shows their observations, together with a cartoon illustrating the transfer of matter from a normal star, producing a disk of X-ray-emitting ``accreting'' material about a jet-emitting black-hole.
Astronomers call such systems ``microquasars'', because they are scaled down versions of the ``quasar'' phenomenon in active galaxies containing supermassive black holes. 
Because the time scales in these binary systems are so much shorter, such observations provide the opportunity to study the dynamics of relativistic jets on human time scales.

\chandra\ observations detected first a bright knot of what is believed to be synchrotron emission from first one jet (to the left in the figure), and then an opposing jet (to the right).
A comparison of images at different times yields velocities of about half the speed of light.
Four years after the outburst, the bright knots had moved more than 3 light years apart, with the left knot slowing down and fading.
Astronomers infer the synchrotron nature of the jet emission primarily from the power-law shape of the energy spectrum (Compton radiation is an alternative) extending from the radio through the X-ray bands; however, Compton radiation might also account for some of the emission.
The associated Lorentz factors are of order $2\times10^7$ corresponding to electron energies of 10 TeV under the equipartition assumption which gives a magnetic field of 0.3 mG. 

\subsection{X-Ray jets in Radio Galaxies\label{ss:jets}}

A fraction of all galaxies have jet-like structures that emanate from the nucleus and were first discovered in radio images.
These galaxies are known as ``radio galaxies'' and the radio jets are believed to be the synchrotron emission from plasma flowing out from the nucleus of the galaxy at supersonic speeds, in some cases near the speed of light.
In most of these nearby radio galaxies there are also X-ray emitting jets.
As with the stellar-mass black holes in binary systems (``microquasars'') the jet X-ray emission appears to be synchrotron radiation.
The interest in these systems arises because the X-ray-synchrotron-emitting electrons are at considerably higher energy than those electrons emitting   radiation in the visible and in the radio.
The synchrotron lifetime for these electrons is short compared to the travel time down the jet thus requiring in-situ particle acceleration.  
Often-seen offsets between X-ray-bright and radio-bright regions give clues to the particle acceleration mechanism.

\begin{figure}
\begin{center} 
\epsfysize=6cm
\epsfbox{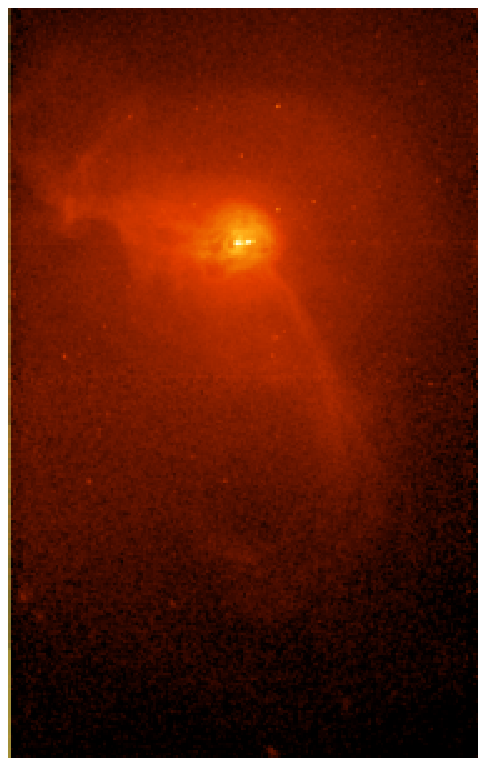} 
\epsfysize=6cm
\epsfbox{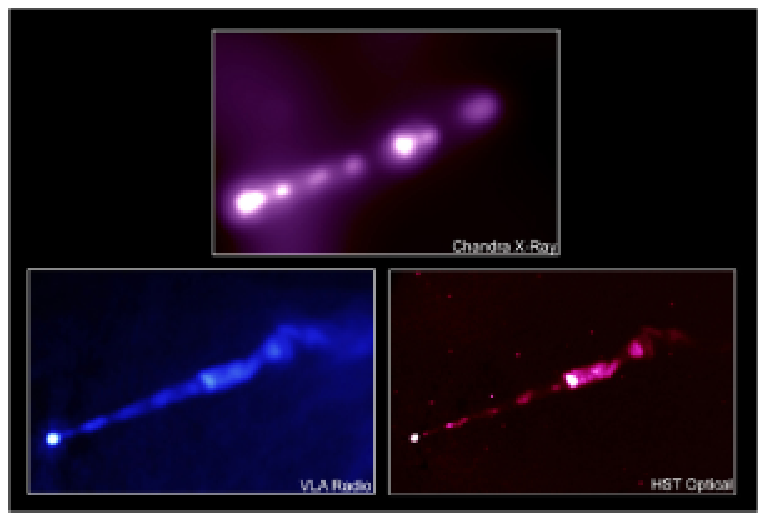} 
\caption{Left: X-ray image of the giant elliptical galaxy M87. The diffuse X-ray emission on the left panel is from hot gas bound by gravity to the central galaxy. The image is 8.6 x 13.7 arcmin. Right: The X-ray synchrotron jet emanating from the nucleus as seen in three wavelength bands. Each panel is 32 x 21 arcsec. At the distance of M87, 1 arcsec is roughly 260 light years. Courtesy NASA.
\label{f:m87}}
\end{center}
\end{figure}

The left portion of Figure~\ref{f:m87} shows the X-ray image one of one of these galaxies, M87 (also a Messier catalog object!).
Bright arcs and dark cavities in the multimillion degree gas of M87 accompany the central jet.
Further away, faint rings with two spectacular plumes extend beyond the jet.
These features are evidence of repetitive outbursts from the vicinity of the central supermassive black hole which impact the entire galaxy.
The right portion of Figure~\ref{f:m87}\ of M87 shows a (multi-wavelength) closeup of the central jet. 
The \chandra\ X-ray image (Marshall et al. 2002) displays an irregular, knotty
structure similar to that seen at radio and optical (Perlman et al. 2001) wavelengths. 
Note that the knots near the central core are much brighter in X rays.
The X-ray spectra of the two brightest knots are consistent with a simple power-law model and the spectral flux distributions over broader wavelength range are consistent with synchrotron models 

\section{Conclusions and Acknowledgments}

Designed for three years of operation, the \chandra\ X-ray Observatory has now been operating for more than seven years.
The Observatory's capability for high-resolution imaging has enabled detailed
studies of the structure of extended X-ray sources, including supernova remnants, astrophysical jets, and hot gas in galaxies and clusters of galaxies. 
In addition to mapping the structure of extended sources, the superb angular resolution permits studies of discrete sources which would otherwise be impossible.
From planetary systems to deep surveys of the faintest and most distant sources,
the scientific results from \chandra\ operation have been outstanding. 
Equally important have been \chandra's unique contributions to high-resolution
dispersive spectroscopy. 
The high spectral resolution of the \chandra\ gratings isolate individual lines from the myriad of spectral lines, which would overlap at lower resolution.
Through all these observations, users have addressed, and continue to address, the most exciting topics in contemporary high-energy astrophysics.

I acknowledge the major contributions to the success of the Observatory made by the scientists and engineers associated with the instrument teams, the NASA Project at Marshall Space Flight Ceneter, the many contractors, and the Chandra X-Ray Center, with special thanks to its Director, Dr. H. Tananbaum. I gratefully acknowledge conversations with Drs. D. Worrall, H. Marshall, and S. O'Dell who kindly made comments to a draft of this article.
Finally, I wish to acknowledge the tremendous contributions of the now-deceased Telescope Scientist, Dr. Leon Van Speybroeck.


\begin{references}

Corbel, S., Fender, R. P., Tzioumis, A. K., Tomsick, J. A., Orosz, J. A., Miller, J. M., Wijnands, R., \&  Kaaret, P. ``Large-Scale, Decelerating, Relativistic X-ray Jets from the Microquasar XTE J1550-564'' Science, Volume 298, Issue 5591, pp. 196-199 (2002)

Hester, J. J., Mori, K., Burrows, D., Gallagher, J. S., Graham, J. R., Halverson, M., Kader, A., Michel, F. C., \& Scowen, P. ``Hubble Space Telescope and Chandra Monitoring of the Crab Synchrotron Nebula'' The Astrophysical Journal, Volume 577, Issue 1, pp. L49-L52. (2002)

Manchester, R.N. \& Taylor, J.H. "Pulsars", W.H.Freeman and Company, SanFrancisco (1977)

Marshall, H.L., Miller, B.P., Davis, D.S., Perlman, E.S., Wise, M., Canizares,
C.R., \& Harris, D.E. ``High-Resolution X-Ray Image of the Jet in M87'', The Astrophysical Journal, Volume 564, Issue 2, pp. 683-687. (2002)

Perlman, E. S.,  Sparks, W. B., Radomski, J., Packham, C., Fisher, R. S., Piña, R., \& Biretta, J. A. ``Deep 10 Micron Imaging of M87'', The Astrophysical Journal, Volume 561, Issue 1, pp. L51-L54. (2001)

Weisskopf, M. C., Hester, J. J., Tennant, A. F., Elsner, R. F., Schulz, N. S.,
Marshall, H. L., Karovska, M., Nichols, J. S., Swartz, D. A.,
Kolodziejczak, J. J., \& O'Dell, S. L. ``Discovery of Spatial and Spectral Structure in the X-Ray Emission from the Crab Nebula'', The Astrophysical Journal, Volume 536, Issue 2, pp. L81-L84. (2000)

Weisskopf, M. C., Aldcroft, T. L., Bautz, M., Cameron, R. A., Dewey, D., Drake, J. J., Grant, C. E., Marshall, H. L., \& Murray, S. S. ``An Overview of the Performance of the \chandra\ X-Ray Observatory'', Experimental Astronomy, Volume 16, Issue 1, pp. 1-68 (2003).

\end{references}
\end{document}